\documentclass[prl,twocolumn,amsmath,superscriptaddress,amssymb]{revtex4-2}

\usepackage{graphicx}
\usepackage{dcolumn}
\usepackage{bm}
\usepackage{amssymb}
\usepackage{amsmath}
\usepackage{wasysym}
\usepackage{color}
\usepackage{times}

\usepackage{placeins}
\usepackage{epstopdf}

\usepackage[separate-uncertainty=true, multi-part-units=single]{siunitx}

\usepackage{ulem}
\usepackage{nicefrac}
\usepackage[dvipsnames]{xcolor}

\usepackage[colorlinks]{hyperref} 
\hypersetup{breaklinks=true, citecolor=blue, linkcolor=blue, menucolor=blue, urlcolor=blue}

\usepackage{comment}
\usepackage{textcase}

\usepackage[capitalise]{cleveref}

\newcommand{\bra}[1]{\left\langle\,#1\,\right|}
\newcommand{\ket}[1]{\left|\,#1\,\right\rangle}

\makeatletter
\renewcommand*\env@matrix[1][c]{\hskip -\arraycolsep
  \let\@ifnextchar\new@ifnextchar
  \array{*\c@MaxMatrixCols #1}}
\makeatother
\begin{document}

\title{Strongly Anisotropic Spin and Orbital Rashba Effect at a Tellurium~--~Noble Metal Interface}

\author{B. Geldiyev}
\author{M. \"{U}nzelmann}
\email{muenzelmann@physik.uni-wuerzburg.de}
\affiliation{Experimentelle Physik VII and W\"{u}rzburg-Dresden Cluster of Excellence ct.qmat, Universit\"{a}t W\"{u}rzburg, Am Hubland, D-97074 W\"{u}rzburg, Germany}

\author{P. Eck}
\affiliation{Institut f\"{u}r Theoretische Physik und Astrophysik and W\"{u}rzburg-Dresden Cluster of Excellence ct.qmat, Universit\"{a}t W\"{u}rzburg, Am Hubland, D-97074 W\"{u}rzburg, Germany}

\author{T. Ki{\ss}linger}
\affiliation{Lehrstuhl f\"{u}r Festk\"{o}rperphysik, Universit\"{a}t Erlangen-N\"{u}rnberg, Staudtstra{\ss}e 7, D-91058 Erlangen, Germany}

\author{J. Schusser}
\author{T. Figgemeier}
\author{P. Kagerer}
\author{N. Tezak}
\affiliation{Experimentelle Physik VII and W\"{u}rzburg-Dresden Cluster of Excellence ct.qmat, Universit\"{a}t W\"{u}rzburg, Am Hubland, D-97074 W\"{u}rzburg, Germany}

\author{M. Krivenkov}
\author{A. Varykhalov}
\affiliation{Helmholtz-Zentrum Berlin für Materialien und Energie, BESSY II, Albert-Einstein-Stra{\ss}e 15, D-12489 Berlin, Germany}

\author{A. Fedorov}
\affiliation{Helmholtz-Zentrum Berlin für Materialien und Energie, BESSY II, Albert-Einstein-Stra{\ss}e 15, D-12489 Berlin, Germany}
\affiliation{Leibniz Institute for Solid State and Materials Research, IFW Dresden and W\"{u}rzburg-Dresden Cluster of Excellence ct.qmat, Helmholtzstr. 20, D-01069 Dresden, Germany}
\affiliation{Joint Laboratory ``Functional Quantum Materials'' at BESSY II, Albert-Einstein-Stra{\ss}e 15, D-12489 Berlin, Germany}

\author{L. Nicola\"{i}}
\author{J. Min\'{a}r}
\affiliation{New Technologies - Research Center, University of West Bohemia, 30100 Pilsen, Czech Republic}

\author{K. Miyamoto}
\author{T. Okuda}
\author{K. Shimada}
\affiliation{Hiroshima Synchrotron Radiation Center, Hiroshima University, Higashi-Hiroshima 739-0046, Japan}

\author{D. Di Sante}
\affiliation{Department of Physics and Astronomy, University of Bologna, 40126 Bologna, Italy}
\author{G. Sangiovanni}
\affiliation{Institut f\"{u}r Theoretische Physik und Astrophysik and W\"{u}rzburg-Dresden Cluster of Excellence ct.qmat, Universit\"{a}t W\"{u}rzburg, Am Hubland, D-97074 W\"{u}rzburg, Germany}

\author{L. Hammer}
\author{M. A. Schneider}
\affiliation{Lehrstuhl f\"{u}r Festk\"{o}rperphysik, Universit\"{a}t Erlangen-N\"{u}rnberg, Staudtstra{\ss}e 7, D-91058 Erlangen, Germany}

\author{H. Bentmann}
\altaffiliation{Present address: Center for Quantum Spintronics (QuSpin), NTNU Trondheim, NO-7034 Trondheim, Norway}
\author{F. Reinert}
\affiliation{Experimentelle Physik VII and W\"{u}rzburg-Dresden Cluster of Excellence ct.qmat, Universit\"{a}t W\"{u}rzburg, Am Hubland, D-97074 W\"{u}rzburg, Germany}

\date{\today}

\begin{abstract}
We study the interplay of lattice, spin and orbital degrees of freedom in a two-dimensional model system: a flat square lattice of $\mathrm{Te}$ atoms on a $\mathrm{Au(100)}$ surface. The atomic structure of the Te monolayer is determined by scanning tunneling microscopy (STM) and quantitative low-energy electron diffraction (LEED-IV). Using spin- and angle-resolved photoelectron spectroscopy (ARPES) and density functional theory (DFT), we observe a Te-Au interface state with highly anisotropic Rashba-type spin-orbit splitting at the $\bar{\mathrm{X}}$ point of the Brillouin zone. Based on a profound symmetry and tight-binding analysis, we show how in-plane square lattice symmetry and broken inversion symmetry at the Te-Au interface together enforce a remarkably anisotropic orbital Rashba effect which strongly modulates the spin splitting.
\end{abstract}

\maketitle

\begin{figure}[t!]
	\centering
	\includegraphics[width=\columnwidth]{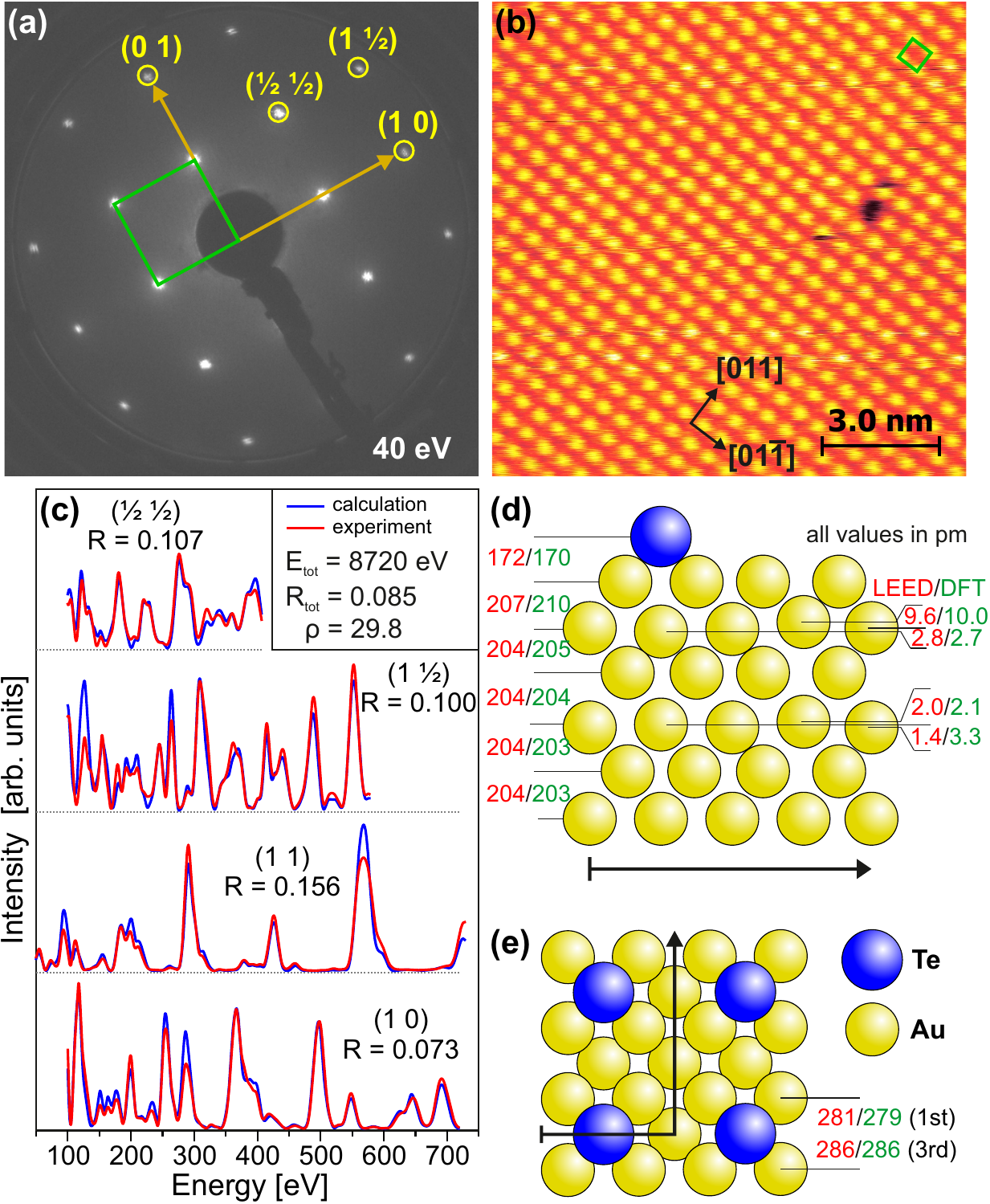}
	\caption{Structural data for the $(2 \times 2)$ phase of $\mathrm{Te}$ on $\mathrm{Au(100)}$. (a) LEED pattern at $40\,\mathrm{eV}$ electron energy. Golden arrows mark the $(1 \times 1)$ unit cell of the unreconstructed $\mathrm{Au(100)}$ surface and the green square the $(2 \times 2)$ Te superstructure cell. (b) STM image ($U = -0.27\,\mathrm{V}$, $I = 0.13\,\mathrm{nA}$ showing the perfectly ordered $(2 \times 2)$ lattice. Te atoms are assigned to the bright protrusions. (c) Selected best-fit LEED-I(V) spectra of the $(2 \times 2)$-Te structure with single beam $\mathrm{R}$-factors close to the overall $\mathrm{R}$-factor value of $\mathrm{R_{\mathrm{P}}} = 0.085$. The fit has a redundancy of $\rho = 29.8$. The complete data set comprises a total energy range of $8720\,\mathrm{eV}$ and is presented in the SM \cite{File_SM}. (d) Ball model of the LEED best-fit structure in side view along the path indicated \textcolor{black}{by the arrow} in (e). (e) Top view. Parameters (\textcolor{black}{atomic distances} in $\mathrm{pm}$) given in (d,e) correspond to the LEED best-fit (red) and the relaxed DFT structure (green) for comparison.}
	\label{fig:Fig0}
\end{figure}

\begin{figure*}[t!]
    \includegraphics[width=\textwidth]{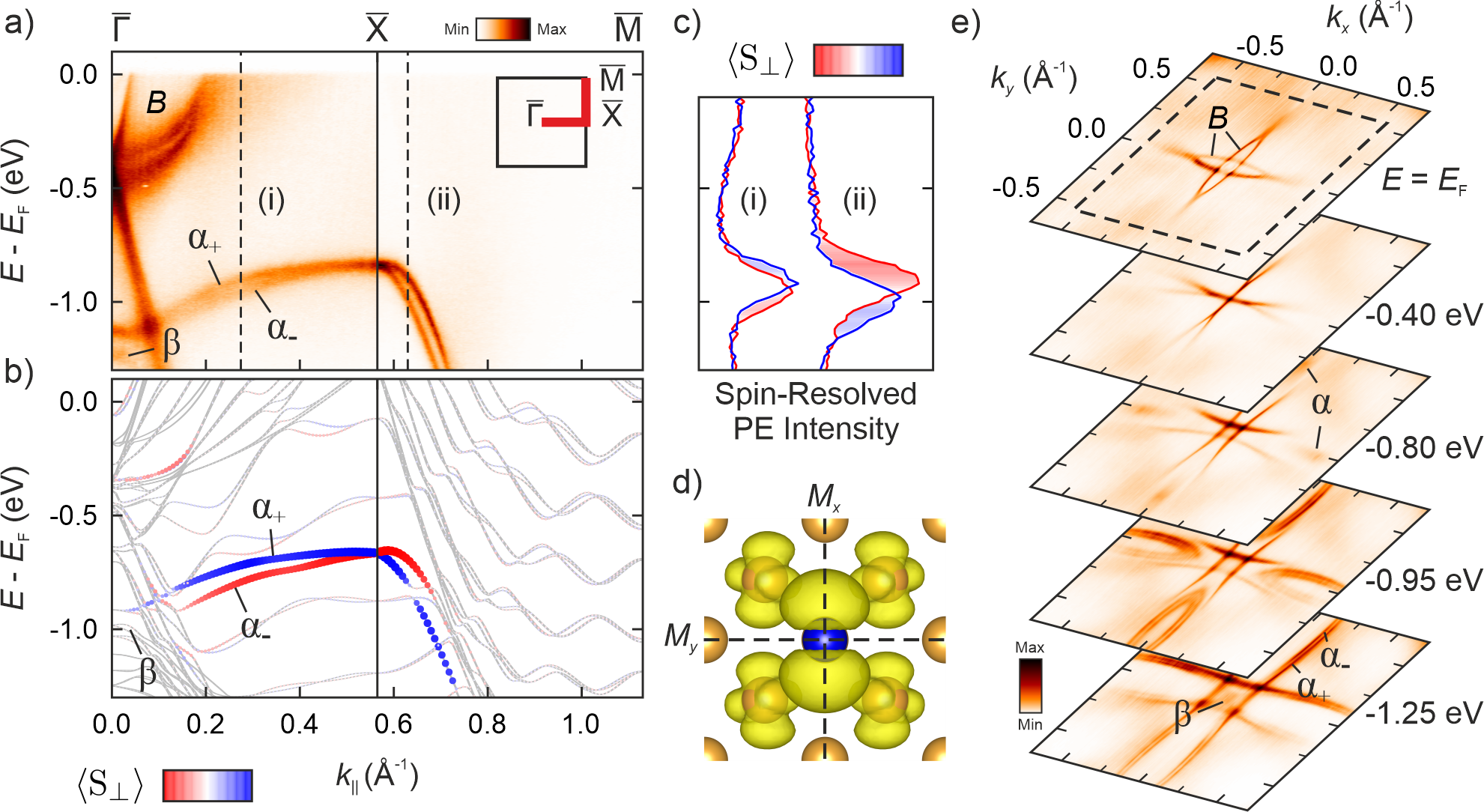}
    \centering
    \caption{Band structure of $\mathrm{Te/Au(100)}$ along (a,b) a $\bar{\Gamma}\bar{\mathrm{X}}\bar{\mathrm{M}}$ path (see inset in (a)) and (e) isoenergy contours at given binding energies. Bands denoted as $\alpha_{\pm}$, $\beta$ and $B$ correspond to adsorbate-induced interface states and substrate bulk bands, respectively. ARPES data in (a) was taken with circularly polarized light ($\mathrm{left} + \mathrm{right}$) at $h\nu = 38\,\mathrm{eV}$. The dot size and red-blue color code in the DFT slab calculation (b) reflect the calculated spin polarization perpendicular to $\mathbf{k_{\parallel}}$. \textcolor{black}{We plotted $\langle S_{\perp} \rangle$ as $\langle S_x \rangle$ along $k_x$ ($\bar{\mathrm{X}}\bar{\Gamma}$) and $-\langle S_y \rangle $ along $k_y$ ($\bar{\mathrm{X}}\bar{\mathrm{M}}$), where blue (red) stands for positive (negative) values, respectively.} (c) Spin-resolved EDC ($h\nu = 38\,\mathrm{eV}$, $p$-polarization) taken at momenta indicated as dashed lines in (a). (d) \textcolor{black}{ DFT-calculated partial charge $|\Psi_{k=\bar{\mathrm{X}}}(\bold{r})|^2$ of $\alpha_{\pm}$ at the $\bar{\mathrm{X}}$ point.}}
	\label{fig:Fig1}
\end{figure*}

\begin{figure}[t!]
	\centering
	\includegraphics[width=\columnwidth]{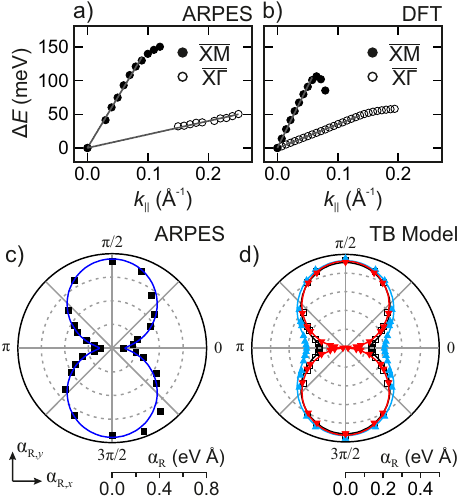}
	\caption{Anisotropy of the Rashba effect around the $C_{2v}$-symmetric $\bar{\mathrm{X}}$ point. (a) Spin splitting extracted from the experimental data (see \cref{fig:Fig1}(a)) along $\bar{\mathrm{X}}\bar{\mathrm{M}}$ and $\bar{\mathrm{X}}\bar{\Gamma}$ lines as a function of $k_{\parallel}$. (b) Same as in (a) but obtained from DFT (see \cref{fig:Fig1}(b)). (c) Rashba parameter $\alpha_{\mathrm{R}}$ as a function of $\phi$ (see text), where $\bar{\mathrm{X}}$ is the origin and $0^{\circ}$ and $90^{\circ}$ being along $\bar{\Gamma}$ and $\bar{\mathrm{M}}$, respectively. (d) Same as in (c) but obtained from the utilized tight-binding model as a function of different $\gamma_2^{\pm 1}$ values (red $\gamma_2^{\pm 1} = 0$, black $\gamma_2^{\pm 1} = 0.2$, and cyan $\gamma_2^{\pm 1} = 0.3$). Markers in (c,d) correspond to data points extracted from the experimental data and TB model, respectively. Solid lines are fits to those (see text).}
    \label{fig:Fig2}
\end{figure}

\begin{figure}[t!]
	\centering
	\includegraphics[width=\columnwidth]{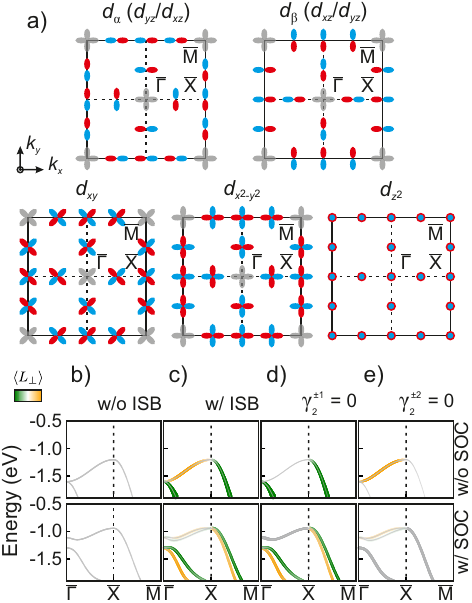}
	\caption{(a) Momentum space textures in a $d$-orbital square lattice. Parity of a respective texture can be identified along specific high-symmetry lines. (b-e) Band structure plots obtained from the tight-binding model as a function of varying ISB parameters $\gamma_{l}^{m_l}$. Top panel in (b-e) is when SOC not being taken into account (color scale adjusted compared to the lower panel for better visualization, \textcolor{black}{see Supplementary Fig.~S8 for direct comparison}). Bottom panel shows the outcome when SOC is considered.}
    \label{fig:Fig3}
\end{figure}

The Rashba effect \cite{Rashba1984, Manchon2015, Bihlmayer2022} gives rise to a momentum-dependent spin splitting in two-dimensional (2D) electron systems \cite{LaShell1996, Nicolay2001, Hoesch2004, Ast2007, Bentmann2009, King2011, Marchenko2012, Sunko2017, Uenzelmann2020} with inversion symmetry breaking (ISB) acting in a polar out-of-plane fashion. The latter causes an alignment of the electrons spin perpendicular to its wave vector via spin-orbit coupling (SOC), similar to topological surface states \cite{Hsieh2009}, paving the way towards \textit{spintronics}.
\\
Anisotropic spin-orbit effects arise when in-plane lattice symmetries are taken into account, such as  valley splittings in systems with $C_{3h}$ symmetry \cite{Bauernfeind2021, Riley2014, Suzuki2014, Eickholt2018, Souma2011, Fu2009}. Moreover, anisotropic Rashba splittings can emerge around time-reversal invariant momenta (TRIM) with reduced point group symmetries such as $C_{2v}$ or $C_{1h}$ \cite{Oguchi2009, Simon2010, Vajna2012, Miyamoto2012, Sakamoto2016}. Those effects are usually treated in the framework of $\mathbf{k \cdot p}$ theory \cite{Oguchi2009, Simon2010, Fu2009, Vajna2012} directly incorporating real-space point group symmetries into the momentum-space electronic structure. However, this provides only limited insights into the underlying microscopic mechanisms.
\\
Recently, the so-called orbital Rashba effect (ORE) \cite{Park2012B, Go2021_Cu}
\noindent
\begin{equation}
    H_{\mathrm{ISB}} = \alpha_{\mathrm{ISB}} \cdot \mathbf{L} (\mathbf{z} \times \mathbf{k})~,
    \label{H_ORE}
\end{equation}
\noindent
related to the ISB-induced formation of a chiral orbital angular momentum (OAM) $\mathbf{L}$, has been established as the origin of the spin Rashba effect \cite{Park2011, Park2012B, Uenzelmann2020}. In such manner, control and enhancement of the ISB strength $\alpha_{\mathrm{ISB}}$ further allows to maximize spin-orbit splittings up to the atomic SOC energy \cite{Sunko2017}. The ORE has also been discussed as one of the main driving forces of spintronic phenomena \cite{Go2018, Ding2020, Bhowal2020} and, likewise, orbital precursors were predicted in the context of spin-valley physics \cite{Cysne2021, Bhowal2021}. Overall, this leads to the new field \textit{orbitronics}, exploiting OAM itself as the leading quantum degree of freedom \cite{Bernevig2005, Jo2018, Go2021, Go2020, Canonico2020, Lee2021, Ding2022}.
The ORE, in turn, represents a key element of this approach.
Moreover, due to its connection to Berry curvature \cite{Schueler2019, Lesne2023}, the OAM has proven to be an important observable for the study of topological quantum matter \cite{Cho2018, Uenzelmann2021}. However, in contrast to the spin Rashba effect, the fundamental influence of more complex effects of lattice symmetries on the ORE has remained largely unexplored.
\\
In this work, we establish a mechanism by which lattice-modulated orbital textures enforce a highly anisotropic ORE and spin-orbit splitting.
To this end, we investigate the Rashba effect in a square lattice $\mathrm{Te/Au(100)}$.
Adsorption of $\mathrm{Te}$ on a $\mathrm{Au(100)}$ substrate leads to the formation of interface states with Te-$p$ and Au-$d$ characters. Using experiments and model calculations, we show how the symmetry-selective formation of OAM gives rise to a two-fold anisotropy of the Rashba parameter for these interface states. 
The key ingredient is the momentum-space orbital texture, which is governed by the underlying point group symmetries. These findings deepen the understanding of the microscopic interplay between lattice symmetries and spin-orbit coupling effects.
\\
Symmetry analysis of the electronic states in $\mathrm{Te/Au(100)}$ requires a precise verification of the atomic arrangement. Therefore, we first focus on the analysis of the surface atomic structure employing LEED-I(V), STM as well as DFT, and subsequently discuss the electronic structure based on ARPES experiments, DFT and tight-binding (TB) band structure calculations. Details about the experimental and theoretical methods can be found in the Supplemental Material \cite{File_SM}.
\\
Deposition of $\nicefrac{1}{4}\,\mathrm{ML}$ of $\mathrm{Te}$ (where $1\,\mathrm{ML}$ is defined as the surface atomic density of $\mathrm{Au(100)}$) at $90\,\mathrm{K}$ and post-annealing the sample to $820\,\mathrm{K}$ results in a well-ordered $(2 \times 2)$ phase (see \cref{fig:Fig0}(a)). Atomically resolved STM images (see \cref{fig:Fig0}(b)) reveal that within the unit cell one bright protrusion is observed which must be associated with Te atoms. We note, that our calibration of the Te amount to about $1\,\mathrm{\%}$ of a $\mathrm{ML}$ relies on a series of other structures that we studied previously \cite{Uenzelmann2020, Kisslinger2020, Kisslinger2021}. Based on this, the LEED-I(V) analysis was restricted to four physically plausible models, the Te-Au exchange at the surface, the Te adsorption on top-, bridge- and hollow-sites. The first three could be ruled out in regard to Pendry $\mathrm{R}$-factors $\mathrm{R_{\mathrm{P}}} \geq 0.6$ against the hollow-site model. The LEED analysis is guided by the $\mathrm{R}$-factor that describes the matching between experimental and calculated intensity spectra. Here, $\mathrm{R_{\mathrm{P}}} = 1$ means completely non-correlated and $\mathrm{R_{\mathrm{P}}} = 0$ stands for perfect match. After further refinement of the model parameters we obtain a final best-fit $\mathrm{R_{\mathrm{P}}} = 0.085$ at a data redundancy of $\rho = 29.8$. This high fit quality is visualized by the close-to-perfect match of measured and calculated best-fit LEED-I(V) spectra, which are depicted in \cref{fig:Fig0}(c). Complete set of data is provided in Supplementary Note IV. Further evidence for the correctness of the hollow-site model comes from the close correspondence of derived LEED best-fit parameter values with those obtained for the corresponding relaxed DFT model (within $\pm 3\,\mathrm{pm}$) for every adjusted parameter (see \cref{fig:Fig0}(d,e)).
\\
With the precise knowledge of the atomic structure, we can now evaluate the electronic structure of $\mathrm{Te/Au(100)}$. \cref{fig:Fig1}(a) shows an ARPES measurement taken along a $\bar{\Gamma}\bar{\mathrm{X}}\bar{\mathrm{M}}$ path in the surface Brillouin zone (BZ). A variety of features crossing the Fermi level are found (labeled as $B$), which correspond to bulk bands of the substrate back-folded by the ($2 \times 2$) superstructure. In excellent agreement with our DFT band structure calculation shown in \cref{fig:Fig1}(b), two more bands, named $\alpha_{\pm}$ and $\beta$, can be recognized that arise upon adsorption of the Te layer. While $\beta$ is only visible in a small $k_{\parallel}$ area close to $\bar{\Gamma}$ owing to the strong hybridization with the projected substrate bulk bands, $\alpha_{\pm}$ is clearly visible within the entire $k_{\parallel}$ range. It has a maximum energy of $-0.84\,\mathrm{eV}$ at the $\bar{\mathrm{X}}$ points and from there evolves into oval hole pockets, as can also be seen in the constant energy cuts in \cref{fig:Fig1}(e). Effective masses of the almost parabolic dispersion correspond to $m^{\ast} = -0.16 \pm 0.01\,m_\mathrm{e}$ and $m^{\ast} = -3.38 \pm 0.02\,m_\mathrm{e}$ along $\bar{\mathrm{X}}\bar{\mathrm{M}}$ and $\bar{\mathrm{X}}\bar{\Gamma}$, respectively, revealing a strong anisotropy around $\bar{\mathrm{X}}$. We find that --- along the considered $k_{\parallel}$ path (see inset in \cref{fig:Fig1}(a)) --- the electronic states in $\alpha_{\pm}$ are predominantly built from Te-$p_y$ orbitals. However, also the $d$-orbitals of the subsurface Au layer have a considerable contribution (see Supplementary Note II and Fig.~S1). In \cref{fig:Fig1}(d), the charge density of $\alpha_{\pm}$, as calculated by DFT at the $\bar{\mathrm{X}}$ point, is shown. In addition to a distinct $p_y$ shape around the Te atom, it is directly apparent that the surrounding Au-$d$ orbitals decisively shape the electronic wave function as well. Overall, $\alpha_{\pm}$ can be considered as a Te-Au interface state arising from hybridization between mirror-symmetry-equivalent Te-$p$ and Au-$d$ orbitals.
\\
Since inversion symmetry is broken at surfaces and interfaces, the formation of spin-orbit splittings is in general allowed for states that are localized there. Indeed, we find a substantial splitting into two spin branches $\alpha_{+}$ and $\alpha_{-}$ away from the $\bar{\mathrm{X}}$ point, which denotes a TRIM of the BZ and thus enforces spin degeneracy. The calculated spin polarization (\cref{fig:Fig1}(b)) reveals a Rashba-type spin-momentum locking perpendicular to $k_{\parallel}$, which is nicely reproduced by our spin-resolved ARPES measurements along $\bar{\mathrm{X}}\bar{\Gamma}$ and $\bar{\mathrm{X}}\bar{\mathrm{M}}$ shown in \cref{fig:Fig1}(c).
\\
The strength of Rashba-type spin splittings is quantitatively described by the Rashba parameter $\alpha_{\mathrm{R}}$ given by the common Rashba Hamiltonian $H_{\mathrm{R}} \propto \alpha_{\mathrm{R}} \mathbf{\sigma} \cdot (\mathbf{z} \times \mathbf{k})$. Hence, it directly determines the slope of the $k$-linear energy splitting, which is plotted in \cref{fig:Fig2}(a) and (b) as extracted from ARPES and DFT, respectively. In a certain range around $\bar{\mathrm{X}}$ one clearly finds $\Delta E \propto k$, while for larger wave vectors deviations arise, most likely due to more pronounced hybridization of these states with bulk bands (see \cref{fig:Fig1}(b)). Interestingly, the Rashba parameter is drastically different along the two normal high-symmetry directions: Along $\bar{\mathrm{X}}\bar{\mathrm{M}}$ one finds an experimental value of $\alpha_{\mathrm{R}}^{\bar{\mathrm{X}}\bar{\mathrm{M}}} = 0.74 \pm 0.01\,\mathrm{eV\AA}$, strikingly similar to the surface state in $\mathrm{AgTe/Ag(111)}$ \cite{Uenzelmann2020} and more than two times larger than the paradigmatic Shockley surface state on $\mathrm{Au(111)}$ \cite{Cercellier2006, LaShell1996, Nicolay2001}. In contrast, the Rashba parameter along $\bar{\mathrm{X}}\bar{\Gamma}$ is more than seven times smaller and reads $\alpha_{\mathrm{R}}^{\bar{\mathrm{X}}\bar{\Gamma}} = 0.10 \pm 0.01\,\mathrm{eV\AA}$. In good agreement, the same trend is obtained from DFT (\cref{fig:Fig2}(b)), i.e., here we find $\alpha_{\mathrm{R}}^{\bar{\mathrm{X}}\bar{\mathrm{M}}} \approx 5 \alpha_{\mathrm{R}}^{\bar{\mathrm{X}}\bar{\Gamma}}$.
\\
Taken together, not only the band dispersion, particularly the band width, but moreover the Rashba effect is strongly anisotropic around the $\bar{\mathrm{X}}$ point of the $\mathrm{Te/Au(100)}$ square BZ. To further elaborate on this aspect, we analyzed the dependence of the Rashba parameter on the polar angle $\phi_k$ in more detail. The result is shown in \cref{fig:Fig2}(c) and clearly proves a nearly two-fold symmetric anisotropic behavior of $\alpha_{\mathrm{R}} (\phi_k)$. The anisotropy around $\bar{\mathrm{X}}$ arises from its two-fold point group symmetry, which in general allows for anisotropic spin splittings \cite{Oguchi2009, Simon2010, Vajna2012, Miyamoto2012}. More precisely, using $\mathbf{k \cdot p}$ theory \cite{Oguchi2009, Simon2010}, it was shown that $C_{2v}$ symmetry leads to a Rashba Hamiltonian
\noindent
\begin{equation}
    H_{\mathrm{R}}^{C_{2v}} \propto \alpha_{\mathrm{R,x}} k_x \sigma_y - \alpha_{\mathrm{R,y}} k_y \sigma_x~,
    \label{RashhaC2v}
\end{equation}
\noindent
yielding two directional Rashba parameters $\alpha_{\mathrm{R,x}}$ and $\alpha_{\mathrm{R,y}}$. This results in a characteristic $\phi_k$-dependence $\alpha_{\mathrm{R}} (\phi_k) = \sqrt{(\alpha_{\mathrm{R,x}}^2 + \alpha_{\mathrm{R,y}}^2)/2 + (\alpha_{\mathrm{R,x}}^2 - \alpha_{\mathrm{R,y}}^2)/2 \cdot \cos{2\phi_k}}~$, which fits well to our experimental data in \cref{fig:Fig2}(c) confirming the latter is properly described by the Hamiltonian in \cref{RashhaC2v}.
\\
We now come back to the symmetry of the spatial wave functions and discuss the influence of the orbital texture on the anisotropic spin splitting.
The calculated charge density (\cref{fig:Fig1}(d)) and orbital projections (see Supplementary Note II and Fig.~S1) reveal a Te-$p_y$ and Au-$d_{yz}$-like orbital character at the $\bar{\mathrm{X}}$ point (at $(k_x,k_y) = (\pi/a,0)$). The wave function is thus two-fold symmetric as expected from $C_{2v}$ symmetry. Overall, two mirror operations have to be considered, which along $\bar{\mathrm{X}}\bar{\Gamma}$ and $\bar{\mathrm{X}}\bar{\mathrm{M}}$ yield an odd symmetry with respect to $M_y \colon y \mapsto -y$, and an even character under $M_x \colon x \mapsto -x$, respectively. This texture is further confirmed by light-polarization-dependent ARPES measurements and appropriate photoemission calculations (see Supplementary Note II and Fig.~S2).
To understand the decisive role of orbital symmetry on the electronic structure, we devise a TB model for a 2D square lattice.
Since the $d$-orbital mirror symmetries already cover those of the $p$-orbital subspace, an effective five-band $d$-orbital model is considered \cite{Markovic2019}, as discussed further in Supplementary Note III. Moreover, our DFT calculations reveal a vanishing contribution of the Te-$p$ orbitals to the Rashba effect in $\mathrm{Te/Au(100)}$, further justifying this assumption (see Supplementary Fig.~S3 and S4). Within a TB picture, the Hamiltonian can be divided into three parts $H = H_0 + H_{\mathrm{ISB}} + H_{\mathrm{SOC}}$. The first term describes the unperturbed system and the bands are classified in terms of a characteristic symmetry-induced momentum-space orbital texture, as shown in \cref{fig:Fig3}(a). The bands $\alpha_{\pm}$ and $\beta$ belong to the textures $d_{\alpha}$ and $d_{\beta}$, respectively. Without ISB these bands do not hybridize with those of the $\lbrace d_{xy},d_{x^2-y^2},d_{z^2} \rbrace$ manifold, due to the different out-of-plane symmetry $M_z \colon z \mapsto -z$. At the Te-Au interface, Rashba-type breaking of inversion symmetry, in turn, enables hybridization parameters given in the second term $H_{\mathrm{ISB}}$. This drives the formation of finite OAM expectation values $\langle L \rangle$ in the eigenstates, i.e., for the given ISB term, the ORE (\cref{H_ORE}).
Atomic SOC ($H_{\mathrm{SOC}} = \lambda_{\mathrm{SOC}} \mathbf{L \cdot S}$) transforms the ORE into the spin Rashba effect by alignment of OAM and spin, resulting in an energy splitting of the OAM carrying bands \cite{Park2011, Park2012B, Uenzelmann2020}.
This effect can be seen directly in the model calculations shown in \cref{fig:Fig3}(b,c): In the inversion symmetric case (\cref{fig:Fig3}(b)), OAM is quenched in the eigenstates, which leads to a complete suppression of the spin splitting. In contrast, breaking inversion symmetry (\cref{fig:Fig3}(c)) leads to the formation of OAM and Rashba splittings occur upon inclusion of SOC (lower panel).
\textcolor{black}{
In the latter case, the OAM is oriented anti-parallel in the two spin-split branches of a given band. This is evident not only in our toy model calculation but also in the full DFT-based Wannier model (see Fig.~S4). These observations show that in Te/Au(100) the atomic SOC strength is larger than the energy scale associated with ISB, $E_\mathrm{ISB}=\langle H_\mathrm{ISB}\rangle$. 
A more detailed discussion of this aspect can be found in Supplementary Note III.
It should also be noted that the absolute OAM in $\alpha_\pm$ is weaker along ${\bar{\mathrm{X}}\bar{\Gamma}}$ compared to ${\bar{\mathrm{X}}\bar{\mathrm{M}}}$. This results from the fact that the eigenstates along the two high symmetry directions correspond to different OAM manifolds. That is, the maximum expectation values are $|L_x|\leq 2\,\hbar$ along ${\bar{\mathrm{X}}\bar{\mathrm{M}}}$ but only $|L_y|\leq 1\,\hbar$ along ${\bar{\mathrm{X}}\bar{\Gamma}}$. This aspect will be particularly crucial for the following discussion.}
\\
Similar to the experimental observation, one finds this splitting to be anisotropic, as shown in the polar plot (black markers and line) in \cref{fig:Fig2}(d). \textcolor{black}{The absolute magnitude of $\alpha_\mathrm{R}$ differs from \cref{fig:Fig2}(c) due to the semi-quantitative nature of the TB model.}
To examine the anisotropy in more detail (see also Supplementary Note III for further discussion), we will again focus on the ISB term $H_{\mathrm{ISB}}$. Essentially, two separate hybridization terms $\gamma_l^{m_l}$ emerge, namely $\gamma_2^{\pm 1} \propto \bra{d_{xz,yz}} H \ket{d_{xy}}$ and $\gamma_2^{\pm 2} \propto \bra{d_{xz,yz}} H \ket{d_{z^2,x^2-y^2}}$. In this notation, $m_l$ refers to the magnetic quantum number, where the quantization axis is in-plane and perpendicular to $\mathbf{k_{\parallel}}$, i.e., the OAM has a chiral Rashba-like vortex texture. With the OAM operators $\mathbf{L}^{m_l}$ of the corresponding orbital subspace, $H_{\mathrm{ISB}}$ can be decomposed into two contributions, arising from the distinct hybridization terms. We can now explore the effect of individual contributions on the formation of OAM. The crucial aspect here is that the in-plane mirror symmetry of the respective orbitals in the texture (\cref{fig:Fig3}(a)) must be taken into account, since $\gamma_2^{\pm 1}$ and $\gamma_2^{\pm 2}$ couple odd and even states, respectively.
Accordingly, in the calculation in \cref{fig:Fig3}(d) without SOC (upper panel), we find that $\gamma_2^{\pm 2}$ forms OAM in $\alpha$ along $\bar{\mathrm{X}}\bar{\mathrm{M}}$, while there is no OAM along $\bar{\mathrm{X}}\bar{\Gamma}$. In contrast, for $\gamma_2^{\pm 1}$ (\cref{fig:Fig3}(e)), the opposite behavior is observed, i.e., OAM appears only along $\bar{\mathrm{X}}\bar{\Gamma}$. Notably, this anisotropy consistently reverses for the $d_{\beta}$ band due to the inverted orbital texture.
\\
In direct analogy to \cref{RashhaC2v}, we can express the \textit{anisotropic orbital Rashba effect} (see Supplementary Note III) for $\alpha$ around the $C_{2v}$-symmetric $\bar{\mathrm{X}}$ point ($(\pi/a,0)$) as
\noindent
\begin{equation}
    H_{\mathrm{ISB}}^{d,\alpha} (\bar{\mathrm{X}}) = 
    \alpha_{\mathrm{ISB}}^{(2)} \cdot L_x^{(2)} k_y - \alpha_{\mathrm{ISB}}^{(1)} \cdot L_y^{(1)} k_x~,
    \label{H_ORE_ani}
\end{equation}
\noindent
where $\alpha_{\mathrm{ISB}}^{(1)} = -\gamma_2^{\pm 1} a$ and $\alpha_{\mathrm{ISB}}^{(2)} = \gamma_2^{\pm 2} a$ are the directional orbital Rashba parameters with $a$ being the lattice constant.
\\
Including SOC in \cref{fig:Fig3}(d,e) (lower panel), we find that the anisotropic ORE converts into the anisotropic spin Rashba effect (see \cref{RashhaC2v}). For $\gamma_2^{\pm 1} = 0$ (\cref{fig:Fig3}(d)), the suppression of OAM in $\alpha$ along $\bar{\mathrm{X}}\bar{\Gamma}$ enforces spin degeneracy along the same $k$-path. Vice versa, for $\gamma_2^{\pm 2} = 0$, we find that spin degeneracy is preserved along $\bar{\mathrm{X}}\bar{\mathrm{M}}$ (\cref{fig:Fig3}(e)).
\\
From the polar plot in \cref{fig:Fig2}(d), it becomes further evident that the splitting anisotropy can be controlled by $\alpha_{\mathrm{ISB}}^{(1,2)}$. By tuning $\alpha_{\mathrm{ISB}}^{(1)}$ to smaller (red) or larger (cyan) values, while keeping $\alpha_{\mathrm{ISB}}^{(2)}$ fixed, the anisotropic Rashba effect changes accordingly. 
\textcolor{black}{The Rashba parameters $\alpha_\mathrm{R,(x,y)}$ are proportional to the atomic SOC strength $\lambda_\mathrm{SOC}$ and to the orbital Rashba parameters, which quantify the energy scale of SOC and ISB, respectively. The anisotropy is determined by $\alpha_\mathrm{ISB}^{(1,2)}$:}
\noindent
\begin{equation}
    \alpha_{\mathrm{R,x}} \propto \lambda_\mathrm{SOC} \cdot \alpha_{\mathrm{ISB}}^{(1)}~;
    \hspace{0.6cm}
    \alpha_{\mathrm{R,y}} \propto \lambda_{\mathrm{SOC}} \cdot \alpha_{\mathrm{ISB}}^{(2)}~.
\end{equation}
\noindent
\\
That is, we can attribute the experimentally observed anisotropy of the spin-orbit splittings in $\mathrm{Te/Au(100)}$ to a momentum-selective formation of OAM, the so-called anisotropic ORE (\cref{H_ORE_ani}). A key ingredient for this effect is the orbital texture (see \cref{fig:Fig3}(a)), which is dictated by the $C_{2v}$ symmetry of the $\bar{\mathrm{X}}$ points in the square BZ. Using ARPES measurements with variable linear light polarization (see Supplementary Note II and Fig.~S2), we find direct experimental signatures of the crucial underlying orbital symmetries.
We have shown that the ORE and the OAM-based paradigm of spin-orbit splittings --- which have recently received broad attention \cite{Sunko2017, Markovic2019, Uenzelmann2020, Sakamoto2020, Uenzelmann2021, Bauernfeind2021} --- are generally applicable to distinct point group symmetries like $C_{2v}$. As such, our findings will be crucial for future investigations of \textit{orbitronics} phenomena, such as orbital torques \cite{Go2020, Lee2021}, the orbital Hall- and \cite{Go2018} Rashba-Edelstein effect \cite{Ding2022}.
\\
Controlling momentum-dependent spin splittings constitutes an important task for the exploration of novel spin-based quantum materials. In this respect, particularly ISB has been established as one of the major tools \cite{Sunko2017}. The symmetry-selective formation of OAM and resulting anisotropies in Rashba splitting pave the way toward spin and orbital engineering in 2D materials. This provides prospects for the design of optimized spintronics materials exploiting symmetry effects and orbital hybridization.

\section{Acknowledgments}

This work has been supported by the Deutsche Forschungsgemeinschaft (DFG) through the Würzburg-Dresden Cluster of Excellence \textit{ct.qmat} (EXC 2147, Project ID 390858490) and SFB1170 'ToCoTronics' (Projects A01 and C05), Project RE1469-13-2 at JMU Würzburg and Project 497265814 at FAU Erlangen-Nürnberg. D.S. has received funding from the European Union’s Horizon 2020 research and innovation programme under the Marie Skłodowska-Curie Grant Agreement No. 897276. The spin-ARPES measurements were performed with the approval of the Proposal Assessing Committee of the Hiroshima Synchrotron Radiation Center (Proposal No.~22AG034). J.~M.~and L.~N.~also acknowledge support from the CEDAMNF (Grant No.~CZ.02.1.01/0.0/0.0/15\_003/0000358) of the Ministry of Education, Youth and Sports (Czech Republic).



\end{document}